\documentclass[journal=jctc,manuscript=article,layout=twocolumn]{achemso}

\usepackage{graphicx}
\usepackage{amsmath}
\usepackage{amssymb}
\usepackage{color}
\usepackage{rotating}
\SectionNumbersOn

\title{The interaction-strength interpolation method for main-group chemistry: benchmarking, limitations, and perspectives}
\author{Eduardo Fabiano}
\affiliation{Istituto Nanoscienze-CNR, Euromediterranean Center for Nanomaterial Modelling and Technology (ECMT), Via per Arnesano 16, 73100 Lecce, Italy.}
\alsoaffiliation{Istituto Italiano di Tecnologia (IIT), Center for Biomolecular Nanotechnologies@UNILE, Via Barsanti, 73010 Arnesano, Italy.}
\email{eduardo.fabiano@nano.cnr.it}
\author{Paola Gori-Giorgi}
\affiliation{Department of Theoretical Chemistry and Amsterdam Center for Multiscale Modeling, FEW, Vrije Universiteit,
De Boelelaan 1083, 1081HV Amsterdam, The Netherlands}
\author{Michael Seidl}
\affiliation{Department of Theoretical Chemistry and Amsterdam Center for Multiscale Modeling, FEW, Vrije Universiteit,
De Boelelaan 1083, 1081HV Amsterdam, The Netherlands}
\author{Fabio Della Sala}
\affiliation{Istituto Nanoscienze-CNR, Euromediterranean Center for Nanomaterial Modelling and Technology (ECMT), Via per Arnesano 16, 73100 Lecce, Italy.}
\alsoaffiliation{Istituto Italiano di Tecnologia (IIT), Center for Biomolecular Nanotechnologies@UNILE, Via Barsanti, 73010 Arnesano, Italy.}
\date{\today}

\begin{document}

\begin{abstract}
We have tested the original interaction-strength-interpolation (ISI) 
exchange-correlation functional for main group chemistry.
The ISI functional is based on an interpolation between the weak 
and strong coupling limits and includes exact-exchange 
as well as the G\"orling-Levy second-order energy. 
We have analyzed in detail the basis-set dependence of the ISI functional, 
its dependence {on} the ground-state orbitals, 
and the influence of the size-consistency {problem}. 
We show and explain some of the expected limitations of the ISI functional 
(i.e. for atomization energies), but also unexpected results, 
such as the good performance for the interaction energy of 
dispersion-bonded complexes {when the ISI correlation is used as a correction to Hartree-Fock}. 
\end{abstract}

\maketitle

\section{Introduction}
Current approximations for the exchange-correlation functional of Kohn-Sham (KS) density functional theory (DFT) work for systems that 
are weakly or moderately correlated, as they are based on information (exact or approximate) from the weakly correlated regime, when
 the physical system is not too different from the KS one.
The idea of including information from the opposite limit of infinite correlation dates back to Wigner,\cite{Wig-PR-34,Wig-TFS-38} 
who approximated the correlation energy of the uniform electron gas by interpolating between the limits of zero and infinite
 interaction strength. Seidl and coworkers\cite{SeiPerLev-PRA-99,SeiPerKur-PRL-00} imported this idea in the framework of KS DFT. 
They analyzed the structure\cite{Sei-PRA-99,SeiPerLev-PRA-99} of the DFT limit of infinite coupling strength, proposed a 
semilocal approximation for it,\cite{SeiPerKur-PRA-00} and built an exchange-correlation (xc) functional by interpolating along 
the adiabatic connection between zero and infinite interaction strength (``interaction-strength interpolation,'' or ISI). 
The original ISI functional interpolates between exact ingredients at weak coupling (exact exchange and  second-order perturbation theory)
 and approximate ingredients at infinite coupling strength, given by the semilocal ``point-charge plus continuum'' (PC) model.\cite{SeiPerKur-PRA-00,SeiPerKur-PRL-00}

In the recent years, the exact solution for the limit of infinite interaction strength in DFT has 
been derived:\cite{SeiGorSav-PRA-07,GorVigSei-JCTC-09} it is given by a highly non-local functional of the density, 
and can be mapped into a mathematical problem appearing in mass-transportation theory.\cite{ButDepGor-PRA-12,CotFriKlu-CPAM-13} 
Comparison against these exact results showed that the PC model (with a minor readjustment on the next leading term\cite{GorVigSei-JCTC-09}) 
is a rather accurate approximation for the xc energy at infinite coupling strength,\cite{SeiGorSav-PRA-07,GorVigSei-JCTC-09}  while 
its functional derivative misses the non-local features of this limit needed to describe many strong-correlation phenomena in 
DFT in a {\em spin restricted} framework.\cite{MalGor-PRL-12,MirSeiGor-PRL-13,MenMalGor-PRB-14} 
Another approximation for the strong-coupling limit that retains some of its non-locality (the ``non-local radius'' model, or NLR)
 has been recently proposed in Ref.~\cite{WagGor-PRA-14}, and used by Zhou, Bahmann, and Ernzerhof\cite{ZhoBahErn-JCP-15} to construct
 new xc functionals that use the information at infinite coupling strength.

A formal drawback of the original ISI functional is that it is size consistent
only when a system dissociates into equal fragments.
{This problem is shared by different non-local methods in DFT
(see e.g. Refs. \cite{OdaCap-PRA-09,karolewski13}) and in particular by the} 
approximations based on a global interpolation (i.e., performed on quantities integrated over all space) along 
the adiabatic connection, like the one of
Ref.~\cite{MorCohYan-JCPa-06}. {For the latter,} a possible way to restore size-consistency 
in the usual DFT sense\cite{GorSav-JPCS-08,Sav-CP-09} is to turn to models based on {\em local} interpolations, performed 
in each point of space,\cite{MirSeiGor-JCTC-12} a route that is being presently explored by
 different authors.\cite{ZhoBahErn-JCP-15,VucIroSavTeaGor-JCTC-16,KonPro-JCTC-16,Bec-JCP-13a,Bec-JCP-13b,Bec-JCP-13c} 
An efficient implementation of the ingredients needed for a local interpolation along the adiabatic connection\cite{VucIroSavTeaGor-JCTC-16} 
in the ISI spirit is not yet available, and it is the object of ongoing work. 

While a considerable amount of theoretical work on xc functionals that include in an approximate or exact way the strong-interaction 
limit has been done, benchmarking has been restricted so far to atomization energies,\cite{SeiPerKur-PRL-00,ZhoBahErn-JCP-15}
 ionization potentials,\cite{ZhoBahErn-JCP-15} or to simple paradigmatic physical\cite{MalGor-PRL-12,MalMirCreReiGor-PRB-13,MenMalGor-PRB-14} 
and chemical\cite{TeaCorHel-JCP-10,MalMirGieWagGor-PCCP-14,MirUmrMorGor-JCP-14,VucWagMirGor-JCTC-15,VucIroSavTeaGor-JCTC-16} models only. 
Very little is known about the performance of such functionals for bigger systems and for other chemical and physical properties, 
and about technical aspects such as their sensitivity to reference orbitals and their basis set dependence. 

The purpose of the present work is to fill this gap, by starting from a systematic study of the ISI functional in its 
original formulation, for which all the ingredients are readily available. This allows us to start to analyze quantitatively which 
effects are well captured by a functional that includes the strong-coupling limit, together with the practical 
consequences of the size-consistency error for heterolytic dissociation, as well as to examine restricted versus 
unrestricted calculations, and other aspects such as sensitivity to the reference orbitals. 
Our main aim is to provide valuable information for a future generation of functionals based on local interpolations
 along the adiabatic connection that can include the strong-coupling limit without violating size 
consistency.\cite{ZhoBahErn-JCP-15,VucIroSavTeaGor-JCTC-16,KonPro-JCTC-16,Bec-JCP-13a,Bec-JCP-13b,Bec-JCP-13c}

As we shall see, our results show some of the expected limitations of the original ISI functional, but also unexpected results, 
like an excellent performance for the interaction energy of dispersion-bonded complexes that definitely deserves further study. 

\section{Theoretical background}
\label{Sec:theory}
The ISI xc functional\cite{SeiPerLev-PRA-99,SeiPerKur-PRA-00,SeiPerKur-PRL-00} is built by modeling the standard density-fixed 
linear adiabatic connection integrand $W_\lambda[\rho]$,\cite{LanPer-SSC-75} 
\begin{equation}
	W_\lambda[\rho]=\langle\Psi_\lambda[\rho]|\hat{V}_{ee}|\Psi_\lambda[\rho]\rangle-U[\rho],
\end{equation}
where $\Psi_\lambda[\rho]$ is the wavefunction yielding the density $\rho$ and 
minimizing $\langle\Psi|\hat{T}+\lambda\,\hat{V}_{ee}|\Psi\rangle$, and $U[\rho]$ is the Hartree (or Coulomb) energy,
with a functional form $W_\lambda^{\rm ISI}[\rho]$ that has the exact weak- and strong-coupling asymptotic behavior,
\begin{eqnarray}
	W_{\lambda\to 0}[\rho] & = & E_x[\rho]+2\,\lambda\,E_{\rm GL2}[\rho]+...,\label{eq_weak}\\
	W_{\lambda\to \infty}[\rho] & = & W_\infty[\rho]+\frac{W'_\infty[\rho]}{\sqrt{\lambda}}+... \label{eq_strong}.
\end{eqnarray}
 Its final form for the xc energy  is obtained as
 \begin{equation}
 	E_{xc}^{\rm ISI}[\rho]=\int_0^1 W_\lambda^{\rm ISI}[\rho]d\lambda,
 \end{equation}
 and reads
\begin{eqnarray}
\label{eq:isif}
&&E_{xc}^{\rm ISI} = W_\infty + \\
\nonumber
&& \quad + \frac{2X}{Y}\left[\sqrt{1+Y}-1-Z\ln\left(\frac{\sqrt{1+Y}+Z}{1+Z}\right)\right],
\end{eqnarray}
where
\begin{equation}
\label{eq:ISIxyz}
X=\frac{xy^2}{z^2},\;\;\; Y=\frac{x^2y^2}{z^4},\;\;\; Z=\frac{xy^2}{z^3}-1,
\end{equation}
and $x=-4E_{\rm GL2}$, $y=W'_\infty$, and $z=E_x-W_\infty$. The ISI functional is thus based on four 
ingredients: two {come} from the limit of weak interaction of Eq.~\eqref{eq_weak} expressed in terms of 
orbital and orbital energies, namely the exact exchange energy 
\begin{equation}\label{exx_eq}
E_x=-\frac{1}{2}\sum_{i,j}\int d \mathbf{r} \int d\mathbf{r}' \frac{\phi^*_{i}(\mathbf{r})\phi_{j}^*(\mathbf{r}')\phi_{j}(\mathbf{r})\phi_{i}(\mathbf{r}')}{|\mathbf{r}-\mathbf{r}'|}
\end{equation}
and the G\"orling-Levy\cite{GorLev-PRA-94} second-order energy
\begin{eqnarray}
\nonumber
E_{\rm GL2} & = &-\frac{1}{4}\sum_{abij}\frac{|\langle\phi_i \phi_j||\phi_a
  \phi_b \rangle|^{2}}{\epsilon_{a}+\epsilon_{b}-\epsilon_{i}-\epsilon_{j}} -
\\
&&-\sum_{ia}\frac{|\langle \phi_{i}|\hat{v}_{\rm x}^{\rm KS}-\hat{v}_{\text{x}}^{\rm HF}|\phi_{a}\rangle|^{2}}{\epsilon_{a}-\epsilon_{i}},
\label{eq:EGL2}
\end{eqnarray}
{where $\langle \cdot\cdot||\cdot\cdot\rangle$ denotes an antisymmetrized
  two-electron integral;}
two {are derived} from the limit of strong coupling of Eq.~\eqref{eq_strong}: $W_\infty[\rho]$ is the indirect part of the minimum possible expectation value of the electron-electron repulsion in a given density,\cite{SeiGorSav-PRA-07} and $W'_\infty[\rho]$ is the potential energy of coupled zero-point oscillations of localized electrons.\cite{GorVigSei-JCTC-09} They are both highly non-local density functionals that are presently expensive to compute exactly.\cite{SeiGorSav-PRA-07,GorVigSei-JCTC-09,MenLin-PRB-13,CheFriMen-JCTC-14,MenMalGor-PRB-14,VucWagMirGor-JCTC-15} They are well approximated\cite{SeiGorSav-PRA-07,GorVigSei-JCTC-09} by the semilocal PC model,\cite{SeiPerKur-PRA-00} which we use in this work,
\begin{eqnarray}
W_\infty[\rho] & = & \int \left[A\rho(\mathbf{r})^{4/3} + B\frac{|\nabla\rho(\mathbf{r})|^2}{\rho(\mathbf{r})^{4/3}}\right]d\mathbf{r} \\
W'_\infty[\rho] & = & \int \left[C\rho(\mathbf{r})^{3/2} + D\frac{|\nabla\rho(\mathbf{r})|^2}{\rho^{7/6}(\mathbf{r})}\right]d\mathbf{r}.
\label{eq:WprimePC}
\end{eqnarray}
The parameters $A=-1.451$, $B=5.317\times10^{-3}$, and $C=1.535$, are determined by the electrostatics of the PC cell,\cite{SeiPerKur-PRA-00} while 
the parameter $D$ cannot be derived in the same way, and different choices are possible.
 For example, we can fix $D$ by requiring that $W'_\infty[\rho]$ be self-interaction free for the H atom density.\cite{SeiPerKur-PRA-00} 
Another possible choice, which was adopted when the ISI functional was first proposed and tested for atomization energies, is to
 fix $D$ by requiring that $W'_\infty[\rho]$ be exact for the He atom density.\cite{SeiPerKur-PRL-00} 
At the time, however, the exact solution for $W'_\infty[\rho]$ was not available, and the accurate $W'_\infty[\rho]$ for 
He was estimated from a metaGGA functional. Few years later, when the exact $W'_\infty[\rho]$ has been evaluated 
for several atomic densities, it has been found that the metaGGA values were not accurate enough.\cite{GorVigSei-JCTC-09}
 The parameter $D$ has then been changed and fixed by using the exact $W'_\infty[\rho]$ for the He atom.
 This choice, corresponding to  $D=-2.8957\times10^{-2}$, improves significantly the 
agreement between the PC model for $W'_\infty[\rho]$ and the exact values for several 
atomic densities\cite{GorVigSei-JCTC-09} and it is the one we use in this work.

To see how the limits of Eqs.~\eqref{eq_weak}-\eqref{eq_strong} are included in
 the ISI functional of Eq.~\eqref{eq:isif}, we can expand $E_{xc}^{\rm ISI}[\rho]$ in a series for small $E_{\rm GL2}$, 
\begin{equation}
E_{xc}^{\rm ISI}\Big|_{E_{\rm GL2}\to 0}= E_x+E_{\rm GL2}+\frac{4}{3(E_x-W_\infty)}E_{\rm GL2}^2+...
\end{equation}
showing that ISI includes the exact-exchange and recovers second-order perturbation theory. 

The opposite limit of strong correlation is normally signaled by the closing of the energy gap between the highest occupied molecular orbital (HOMO)
and the lowest unoccupied molecular orbital (LUMO), 
which usually makes appear a broken symmetry solution with lower energy. If we do not allow symmetry breaking, 
the gap closes, implying that $E_{\rm GL2}\to -\infty$ and 
\begin{eqnarray}
\label{eq:EcGL2mininf}
&&E_{xc}^{\rm ISI}\Big|_{E_{\rm GL2}\to -\infty} = W_\infty+2 W'_\infty -\\
\nonumber
&&\quad\quad\quad - \frac{2\, {W'}_\infty^2}{E_x-W_\infty}\ln\left(1+\frac{E_x-W_\infty}{W'_\infty}\right).
\end{eqnarray}
The first two terms, $W_\infty[\rho]+2 W'_\infty[\rho]$, give the xc energy in the limit of strong coupling, which is 
the sum of a purely electrostatic indirect part ($W_\infty[\rho]$) and electronic zero point oscillations 
(the factor two in front of $W'_\infty[\rho]$ accounts for the zero point kinetic energy,\cite{GorVigSei-JCTC-09} and comes
 from the integration of the term $\sim \lambda^{-1/2}$ in Eq.~\eqref{eq_strong}). The last term in Eq.~\eqref{eq:EcGL2mininf} 
is dependent on the interpolating function, and can change if we choose different forms (see, e.g., the ones
 of Refs.~\cite{GorVigSei-JCTC-09} and \cite{LiuBur-JCP-09}).

If the four ingredients $E_x$,  $E_{\rm GL2}$, $W_\infty$ and $W_\infty'$ are size consistent, then the ISI xc functional is size 
consistent only when a system dissociates into equal fragments, as it can be easily derived from Eq.~\eqref{eq:isif}.
 A detailed and quantitative analysis of the problem is reported
in Sec.~\ref{sec:sizecons}. 

We should notice, however, that within the less usual restricted framework for open shell fragments, which seems crucial 
to capture strong correlation without introducing artificial magnetic order and it is the present focus of a large theoretical
 effort,\cite{CohMorYan-SCI-08,Bec-JCP-13a,Bec-JCP-13b,Bec-JCP-13c}  size-consistency of the $E_x$ and $E_{\rm GL2}$ is lost,\cite{YanMorCoh-JCP-13}
 and usually $E_{\rm GL2}\to-\infty$ at dissociation.  In this case, the ISI xc functional stays finite and tends to the expression 
of Eq.~(\ref{eq:EcGL2mininf}). In this work we have tested the ISI functional following the standard procedure of allowing
 spin-symmetry breaking (for a very recent review on spin symmetry breaking in DFT see Ref.~\cite{GarScuRuzSunPer-MP-16}),
 and we discuss only briefly paradigmatic calculations (the H$_2$ and N$_2$ dissociation curves) in a spin-restricted formalism. 
It is however clear from Eq.~\eqref{eq:EcGL2mininf} that the ISI xc functional is not able to dissociate a single or multiple 
bond properly in a spin-restricted framework, since Eq.~\eqref{eq:EcGL2mininf} will not provide the right energy in this limit. 
The ISI accuracy in the usual unrestricted KS (or Hartree Fock) formalism are less easy to predict, and its analysis is the main object of this work.

\section{Computational details}
The calculations with the ISI xc functional defined by Eqs.~\eqref{eq:isif}-\eqref{eq:WprimePC} have been performed in a 
post-self-consistent-field (post-SCF)
fashion, using reference orbitals and densities obtained from 
different methods; namely, DFT calculations using the 
Perdew-Burke-Ernzerhof (PBE \cite{PerBurErn-PRL-96}), 
the hybrid PBE (PBE0 \cite{AdaBar-JCP-99,PerErnBur-JCP-96}),
and the hybrid Becke-half-and-half (BHLYP \cite{Bec-JCP-93a,Bec-PRA-88,LeeYanPar-PRB-88})
exchange-correlation functionals, the localized
Hartree-Fock (LHF) effective exact exchange method \cite{DelGor-JCP-01}
and the Hartree-Fock (HF) method.

{In different parts of the paper we consider the ISI
  correlation energy, which is defined, as usual in the DFT framework \cite{GriSchBae-JCP-97},
as $E_c^{ISI}=E_{xc}^{ISI}-E_x$, where $E_{xc}^{ISI}$ and $E_x$ are the
ISI xc energy [Eq. (\ref{eq:isif})] and the exact exchange energy
[Eq. (\ref{exx_eq})], respectively. Note that this definition of the
ISI correlation energy is well justified since the ISI xc functional includes
the full exact exchange \cite{SeiPerKur-PRL-00}.}

Unless otherwise stated, all energies have been extrapolated to the
complete basis set limit as described in 
subsection \ref{cbs_sec},
using data from calculations performed with the Dunning
basis set family cc-pV$n$Z ($n=2,\ldots,6$) 
\cite{Dun-JCP-89,KenDunHar-JCP-92,WooDun-JCP-93,WooDun-JCP-94}.
For spin-polarized systems, an UHF formalism
has been employed in the self-consistent calculations.
All calculations have been performed using a development
version of the TURBOMOLE program package \cite{turbo,turbo_rev}.

To assess the performance of the ISI xc functional in practical applications
we considered the following set of tests:

\textbf{Thermochemistry dataset}. It contains atomization energies 
(AE6 \cite{LynTru-JPCA-03,HauKlo-TCA-12}, G2/97 \cite{CurRagRedPop-JCP-97,HauKlo-JCP-12}), ionization potentials
(IP13 \cite{LynZhaTru-JPCA-03}), electron and proton affinities
(EA13 \cite{LynZhaTru-JPCA-03} and PA12 \cite{GoeGri-JCTC-10}), barrier heights
(BH76 \cite{GoeGri-JCTC-10,GoeGri-PCCP-11,ZhaLynTru-JPCA-04,ZhaGonTru-JPCA-05} and K9 \cite{HauKlo-TCA-12,LynTru-JPCA-03a}), 
and reaction energies (BH76RC \cite{GoeGri-JCTC-10,GoeGri-PCCP-11,ZhaLynTru-JPCA-04,ZhaGonTru-JPCA-05} 
and K9 \cite{HauKlo-TCA-12,LynTru-JPCA-03a}) of small main-group molecules.

\textbf{Non-covalent interactions dataset}. It contains
interaction energies of non-covalent complexes having
hydrogen bond (HB6 \cite{ZhaTru-JCTC-05}), dipole-dipole (DI6 \cite{ZhaTru-JCTC-05}),
charge-transfer (CT7 \cite{ZhaTru-JCTC-05}), dihydrogen-bond (DHB23 \cite{FabConDel-JCTC-14}),
and various (S22 \cite{JurSpoCerHob-PCCP-06,MarBurShe-JCP-11}) character.


\subsection{Basis set dependence}
\label{cbs_sec}
\begin{figure}[t]
\includegraphics[width=0.9\columnwidth]{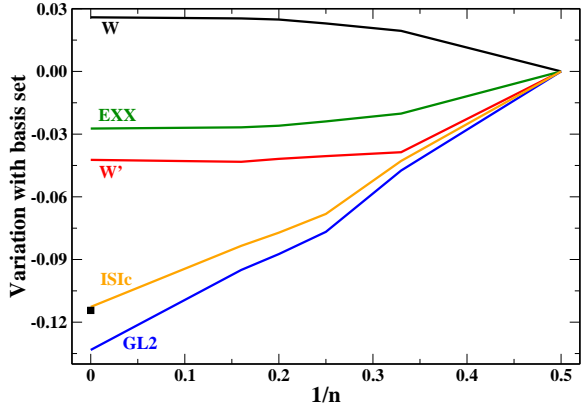}
\caption{\label{basis_set_evolution_fig}Variation with the basis set
  (cc-pV$n$Z) of the various input quantities {(in Ha)} 
used to compute the ISI correlation energy, here for the F atom. The black square at $1/n=0$ indicates the extrapolation obtained applying Eq. (\ref{e1}).}
\end{figure}

The ISI correlation energy formula contains the GL2 correlation
energy of Eq.~\eqref{eq:EGL2}. The latter is well known to exhibit a relevant basis set
dependence as well as a slow convergence to the complete basis set
(CBS) limit. Thus, a similar behavior can be expected also for the
ISI correlation energy. 
Nevertheless, because the ISI energy also includes
other input quantities, whose basis set dependence is different 
from that of GL2, and because all the input quantities
enter non-linearly in the ISI formula, it
is not simple to derive analytically the ISI basis
set dependence. This situation is depicted in Fig.~\ref{basis_set_evolution_fig}, where we report, for the F atom,
the basis set evolution of the different input quantities of the ISI
energy as well as of the ISI correlation energy itself.

\begin{figure}[t]
\includegraphics[width=0.9\columnwidth]{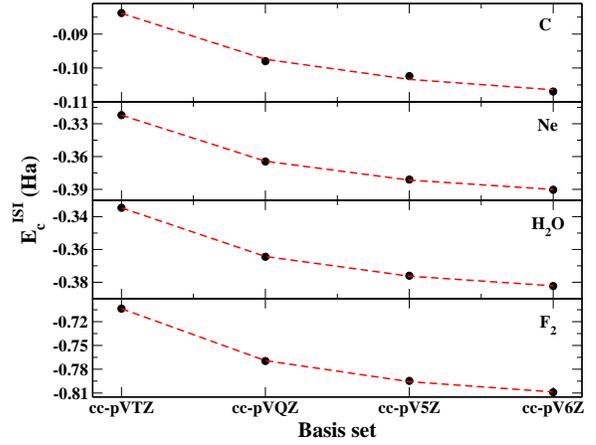}
\caption{\label{fig1}Evolution of the ISI correlation energy of different test
systems with basis set. The red dashed lines denote interpolations obtained
using Eq. (\ref{e1}).}
\end{figure}

For this reason it is not convenient trying to derive the
ISI basis set behavior starting from the assumed behavior
of GL2 and other input quantities, 
as given by popular basis set interpolation-extrapolation formulas
\cite{Mul-BOOK,Tru-CPL-98,Bak-JCP-07,Var-JCP-07,FelPerGra-JCP-11,FabDell-TCA-12}.
Instead, it is more practical to consider the basis set 
evolution of the ISI correlation energy as a whole.
To this end, in analogy with previous works 
on basis set extrapolation
\cite{Tru-CPL-98,Bak-JCP-07,FabDell-TCA-12}, we consider the following ansatz
\begin{equation}\label{e1}
E_c^{\rm ISI}[n] = E_c^{\rm ISI}[\infty] + An^{-\alpha}\ ,
\end{equation}
where the notation $[n]$ indicates that the energy is computed with
an $n$-zeta quality basis set (here specifically the 
cc-pV$n$Z basis set), $A$ is a system-dependent constant,
and $\alpha$ is 
an exponent determining the strength of
the basis set dependence.
Equation (\ref{e1}) provides an accurate fit for the
ISI correlation energies of different systems as we
show in Fig. \ref{fig1} where we report, for some 
example systems, the ISI correlation energies
computed with several basis sets and the corresponding
fit obtained from Eq. (\ref{e1}).
Note also that, as shown in Fig. \ref{basis_set_evolution_fig},
Eq. (\ref{e1}) reproduces correctly the CBS extrapolated value
of the ISI correlation energy as computed using the extrapolated values of
all input ingredients.

Use of Eq. (\ref{e1}) allows to obtain accurate CBS-ISI energies.
However, a more practical approach is to use Eq. (\ref{e1})
into a two-point scheme \cite{Mul-BOOK,Tru-CPL-98,FabDell-TCA-12}, to have the extrapolation formula
\begin{equation}\label{e2}
E_c^{\rm ISI}[\infty] = \frac{E_c^{\rm ISI}[n]n^\alpha - E_c^{\rm ISI}[m]m^\alpha}{n^\alpha - m^\alpha}\ ,
\end{equation}
where $n$ and $m$ label two selected basis sets.
In this work we considered $n=5$ and $m=4$ (for
basis sets smaller than cc-pVQZ we could not
avoid numerical noise in some cases) and fixed the parameter
$\alpha=2.2475$ by fitting to the accurate CBS
ISI correlation energies of atoms He-Ar,
obtained by applying Eq. (\ref{e1}) to the full set
of data corresponding to $n=4,\ldots,6$.
The calculations have been performed in a post-SCF
fashion using LHF orbitals (almost identical results
have been obtained using Hartree-Fock orbitals).
Note that the optimized value of $\alpha$ is 
a bit larger than  the corresponding ones obtained
in Ref. \cite{FasSanTru-JCP-99} for MP2 and CCSD
(1.91 and 1.94, respectively). This indicates that
the ISI correlation converges slightly faster than
the MP2 and CCSD ones to the CBS limit, possibly because it
benefits from the fast convergence of the pure
density-dependent contributions.

\begin{figure}[t]
\includegraphics[width=0.9\columnwidth]{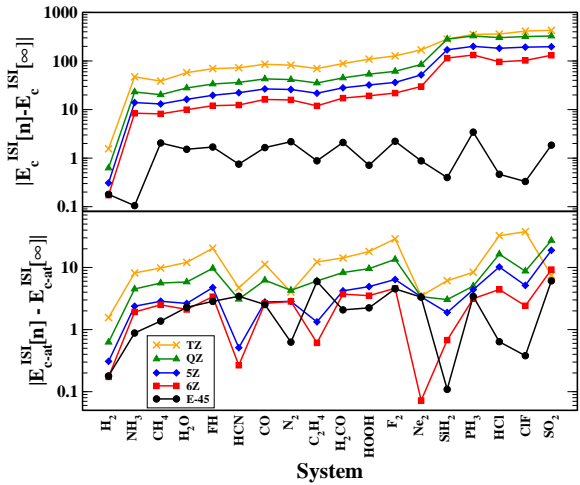}
\caption{\label{fig2} Deviations of the ISI correlation energies
  {(in mHa)} 
computed with various basis sets from the benchmark CBS ones. 
Absolute correlation energies (upper panel) and atomization correlation energies (lower panel).}
\end{figure}

A test of Eq.~\eqref{e2} is reported in Fig.~\ref{fig2}
where we show the errors on ISI absolute correlation energies (upper panel) and atomization correlation energies (lower-panel)
computed with different basis sets, as compared to CBS reference ones, i.e. $E_c^{\rm ISI}[\infty]$ of Eq.~\eqref{e1} fitted to the data with 
$n=3,\ldots,6$ with Eq.~\eqref{e1}.
The results obtained using Eq.~\eqref{e2} are labeled as E-45 in the figure.
%
For the absolute correlation energies we see
that even at the cc-pV6Z level errors of about 10 mHa can be expected,
while only energies obtained via the extrapolation formula
of Eq. (\ref{e2}) show accuracies of about 1 mHa.
For the atomization correlation energies, we deal with energy
differences. {Therefore,} error compensation effects are quite relevant,
especially for the smallest basis sets. Thus, the errors
are close or lower than 10 mHa even for the cc-pVTZ basis set.
Nevertheless, accurate results (about 1 mHa) can be
obtained systematically only using at least a cc-pV5Z basis set
or, even better, via the extrapolation formula of Eq.~\eqref{e2}.

\section{Results}
\label{sec:results}
\subsection{Role of the reference orbitals}
\label{refsect}
The ISI correlation functional is a complicated orbital-dependent
non-linear functional. Thus, a stable self-consistent implementation is a complicated task  
{going} beyond the scope of this paper.  
Here the ISI correlation is employed in a post-SCF scheme,
where the ground-state orbitals and density are computed
using a simpler approach and then used to evaluate the ISI
correlation (and also the exact exchange contribution).

The relevance of the reference density and orbitals for different
DFT calculations has been pointed out 
in several works in literature \cite{JanScu-JCP-08,KimParSonSimBur-JPCL-15}. Therefore,
it appears important to assess the reliability of different
reference orbitals for the calculation of the ISI correlation.
{Furthermore, because the ISI functional is including the GL2
correlation energy as input ingredient, the orbital energies, and in
particular the HOMO-LUMO gap, can be expected to play a major role
(see discussion in subsection \ref{subsec:gap}).}
Hence, we take into account reference ground-state orbitals computed
with the generalized gradient xc approximation of
Perdew-Burke-Ernzerhof (PBE) \cite{PerBurErn-PRL-96}, with the hybrid functionals
PBE0 \cite{AdaBar-JCP-99,PerErnBur-JCP-96} and BH-LYP,\cite{Bec-JCP-93a,Bec-PRA-88,LeeYanPar-PRB-88} 
which include 25\% and 50\% of exact exchange respectively,
with the optimized effective potential named localized Hartree-Fock
(LHF) \cite{DelGor-JCP-01}, and with the Hartree-Fock (HF) method.
{Note that the inclusion of larger fractions of non-local
Hartree-Fock exchange yields increasingly large HOMO-LUMO gaps,
which are also effectively used in double-hybrid functionals \cite{Gri-JCP-06}.}
We remark that the LHF method is {instead} a {\em de facto} exact exchange Kohn-Sham
approach. 
{As such it gives significantly smaller values of the HOMO-LUMO
gap than Hartree-Fock. Moreover,}
it may provide a 
{better} approximation 
{of} the
self-consistent ISI ground-state density and orbitals 
{than approximate functionals or the Hartree-Fock method
(we recall that in general correlation contributions to the density and
orbitals are rather small \cite{jankowski05,fabiano07,grabowski11,Gra-Mol-14}). Anyway,}
we cannot exclude that the 
self-consistent ISI potential {may display non-negligible
differences with respect to the LHF (or the exact exchange) one. 
These differences might concern} {especially a reduction of the HOMO-LUMO gap that will induce a
lowering of the total xc energy (note, however, that for the ISI functional
a complete collapse of the HOMO-LUMO gap is not likely because,
unlike in the GL2 case \cite{RohGriBae-CPL-06}, the large increase of 
kinetic energy associated to it cannot be compensated by the divergence
of the correlation energy, which is bounded from below in ISI)
and cases where static correlation is rather important.}

\begin{figure}[t]
\includegraphics[width=\columnwidth]{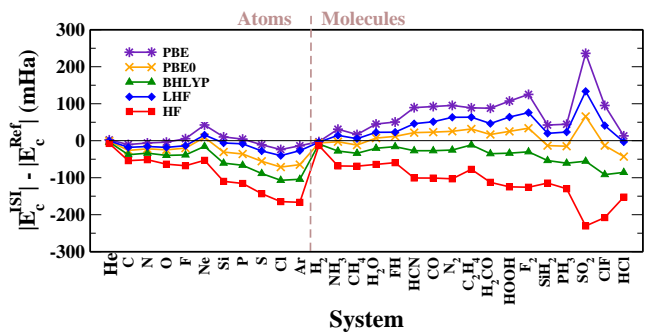}
\caption{\label{fig10}Errors on absolute ISI correlation energies (mHa) calculated using different reference ground-state densities and orbitals.}
\end{figure}

In Table~\ref{tab1} we report the ISI correlation
energies (in absolute value) obtained using different reference ground-state densities and
orbitals (see also Fig. \ref{fig10}). The corresponding GL2 energies are also listed 
in order to compare to the ISI ones (a star is appended to
{the} mean absolute errors reported in Table \ref{tab1} 
to indicate, for each choice of the reference orbitals, the best method between ISI and GL2).
We recall that {the} GL2 and ISI results are extrapolated to
CBS limit, as described in {Eq. (\ref{e2})}
(for GL2 we used Eq. (\ref{e1}) with cc-pVQZ and cc-pV5Z results and the optimized value $\alpha=2.8$).
As already discussed, the expected accuracy of such {an} extrapolation is $\lessapprox 10$ mHa
(this explains the fact that for a few cases, e.g. Si, S, SiH$_2$ using PBE0 orbitals, we have
$|E_c^{\rm ISI}|>|E_c^{\rm GL2}|$, whereas by construction it holds $|E_c^{\rm ISI}| \leq |E_c^{\rm GL2}|$). 
Table~\ref{tab1} also shows {the} correlation energies computed with some
popular semilocal generalized-gradient approximation (GGA) functionals (namely the
Lee-Yang-Parr (LYP) \cite{LeeYanPar-PRB-88}, the PBE, and the PBE with localization (PBEloc)
\cite{ConFabDel-PRB-12} functionals), in order to provide a comparison for the expected accuracy
of standard DFT calculations. 
{The} correlation energies for GGA functionals have been
computed using {the}
cc-pV5Z basis set and {the} PBE self-consistent
orbitals. 
Note that GGA correlation functionals include only dynamical correlation
\cite{Gri-JCP-97,Han-Mol-01}, whereas the ISI method includes both 
dynamical- and static-correlation.

%
\begin{sidewaystable*}
\begin{footnotesize}
\caption{\label{tab1}{\scriptsize Total correlation energies (in mHa, with opposite sign) from semilocal DFT functionals (LYP,PBE, PBEloc, all using PBE orbitals), ISI, and GL2 methods calculated using different reference ground-state orbitals. Reference data are taken from Ref. \cite{ONeGil-MP-05}. The last lines report the mean error (ME) and the mean absolute error (MAE) for each case; a star is appended to the MAEs to indicate, for each choice of the reference orbitals, the best method between ISI and GL2.}}
\begin{tabular}{l|rrr|rrrrr|rrrrr|r}
\hline\hline
Correlation:& LYP & PBE & PBEloc &  \multicolumn{5}{c}{ISI}  & \multicolumn{5}{c}{GL2} & \\ 
\cline{2-4}\cline{5-9}\cline{10-14}
 Orbitals: & PBE & PBE & PBE & PBE & PBE0 & BHLYP & LHF & HF & PBE & PBE0 & BHLYP & LHF & HF & Ref.  \\
\hline
\multicolumn{15}{c}{Closed-shell atoms}\\
He  &  43.7  &  41.1  &  33.8  &  44.3  &  40.9  &  38.3  &  41.8  &  34.5  &  52.3  &  46.9  &  42.8  &  48.4  &  37.4  &  42  \\ 
Ne  &  383.1  &  347.1  &  358.3  &  433.6  &  400  &  375.8  &  406.5  &  338  &  500.3  &  452.9  &  419.8  &  462.6  &  370.1  &  391  \\ 
Ar  &  751.5  &  704.4  &  757.4  &  709.4  &  658.3  &  618.8  &  696.2  &  556.5  &  723.1  &  663.1  &  619.8  &  708  &  558.2  &  723  \\ 
\multicolumn{15}{c}{}\\	
ME  &  7.4  &  -21.1  &  -2.2  &  10.4  &  -18.9  &  -41.0  &  -3.8  &  -75.7  &  39.9  &  2.3  &  -24.5  &  21.0  &  -63.4  &    \\ 
MAE  &  12.7  &  21.1  &  25.1  &  19.5*  &  24.9*  &  41.0*  &  14.2*  &  75.7  &  39.9  &  42.2  &  44.3  &  31.0  &  63.4*  &    \\
\hline
\multicolumn{15}{c}{Open-shell atoms} \\
C  &  158.3  &  144.3  &  139.8  &  145.8  &  129.7  &  118.2  &  138  &  102.1  &  178.6  &  152.4  &  135  &  166.3  &  112.7  &  156  \\ 
N  &  191.9  &  179.9  &  176.4  &  181.9  &  166.2  &  154.4  &  172.5  &  136.6  &  217.3  &  192.9  &  175.6  &  203.2  &  151.3  &  188  \\ 
O  &  256.6  &  235.2  &  234.8  &  251.1  &  230.5  &  215.4  &  237  &  191.7  &  303.2  &  270.4  &  247.7  &  281.4  &  214.2  &  255  \\ 
F  &  321  &  292.6  &  297.4  &  328.4  &  303.1  &  284.6  &  309.6  &  255.4  &  398.4  &  357.7  &  329.5  &  368.7  &  287.6  &  323  \\ 
Si  &  529.2  &  484.2  &  516.8  &  515.3  &  474.5  &  444.4  &  498.9  &  395.2  &  518.3  &  470.3  &  446.3  &  508.8  &  393.5  &  505  \\ 
P  &  566.4  &  526.5  &  564  &  544.4  &  504.1  &  473.6  &  531.6  &  424.5  &  553.3  &  505.5  &  480.6  &  543.3  &  426.7  &  540  \\ 
S  &  627.7  &  584.1  &  626.1  &  592.1  &  547.5  &  514.4  &  575.4  &  459.8  &  600.9  &  546.9  &  517.9  &  585.8  &  456.6  &  603  \\ 
Cl  &  689.7  &  644.5  &  691.6  &  640.2  &  592.8  &  557  &  623.8  &  499.3  &  658.5  &  600.1  &  559.3  &  638.6  &  501.3  &  664  \\ 
\multicolumn{15}{c}{} \\
ME  &  13.4  &  -17.8  &  1.6  &  -4.4  &  -35.7  &  -59.0  &  -18.4  &  -96.2  &  24.3  &  -17.2  &  -42.8  &  7.8  &  -86.3  &    \\ 
MAE  &  13.9  &  17.8  &  20.0  &  9.4*  &  35.7  &  59.0  &  18.4  &  96.2  &  26.2  &  31.0*  &  44.4*  &  18.4  &  86.3*  &    \\ 
\hline 
\multicolumn{15}{c}{Closed-shell molecules} \\
H$_2$  &  38.2  &  42.9  &  37.4  &  38.4  &  35  &  32.2  &  36.9  &  28  &  53.2  &  46.3  &  41.2  &  50.4  &  34.3  &  41  \\ 
NH$_3$  &  318  &  314.2  &  310.8  &  371.2  &  337.2  &  312.1  &  354.7  &  271.9  &  463.3  &  406.7  &  367.6  &  436.3  &  309.4  &  340  \\ 
CH$_4$  &  295  &  300  &  292.5  &  315.8  &  287.3  &  265.1  &  304.5  &  230.2  &  391.5  &  344.9  &  310.8  &  373.5  &  261.1  &  299  \\ 
H$_2$O  &  340.4  &  324.8  &  325.5  &  416.1  &  378.1  &  350.7  &  393.8  &  307  &  514.7  &  452.5  &  410.2  &  478.7  &  347.6  &  371  \\ 
FH  &  362.2  &  335.1  &  340.4  &  439.6  &  400.8  &  373  &  412.1  &  329.6  &  528.9  &  469.1  &  428.2  &  486.8  &  368.2  &  389  \\ 
HCN  &  464.8  &  439.7  &  437.8  &  604.4  &  536.8  &  488  &  561  &  414.3  &  773.8  &  655.2  &  577.1  &  714.7  &  470.1  &  515  \\ 
CO  &  485.2  &  448.4  &  451  &  627.4  &  558.1  &  508  &  586.3  &  434.1  &  787.5  &  670.7  &  593  &  718.2  &  487.9  &  535  \\ 
N$_2$  &  484.2  &  451.5  &  452.5  &  644.5  &  574.5  &  523.8  &  612.4  &  446.1  &  821  &  699.3  &  618.7  &  766.2  &  505.8  &  549  \\ 
C$_2$H$_4$  &  498.7  &  493.7  &  486.8  &  568.9  &  511.7  &  468.6  &  543.6  &  402.7  &  709.3  &  614.6  &  548.4  &  668.3  &  454.8  &  480  \\ 
H$_2$CO  &  540.7  &  514.4  &  514.6  &  673.8  &  602.9  &  550.9  &  632.4  &  473.6  &  844.5  &  725.1  &  644.6  &  775.3  &  534.3  &  586  \\ 
HOOH  &  636.7  &  598.5  &  604.7  &  818.2  &  736.2  &  677.4  &  775.1  &  586.3  &  1023.1  &  885.6  &  794.1  &  951.3  &  663.2  &  711  \\ 
F$_2$  &  675.5  &  612.7  &  627  &  882.2  &  790.8  &  727  &  833  &  631.2  &  1081  &  934.4  &  838.7  &  1003.2  &  705.3  &  757  \\ 
SiH$_2$  &  598.3  &  553.8  &  582.6  &  609.4  &  553.8  &  513.4  &  586.8  &  452.5  &  615.9  &  548.8  &  512.1  &  588.2  &  455.3  &  567  \\ 
PH$_3$  &  676.7  &  642.7  &  677  &  696.6  &  637.1  &  591.4  &  675.6  &  522.3  &  719.9  &  646.1  &  591.4  &  693.8  &  522.8  &  652  \\ 
SO$_2$  &  1257.5  &  1171.3  &  1227.6  &  1570.3  &  1399.7  &  1278.6  &  1467.5  &  1103.5  &  1813.3  &  1559.2  &  1391.1  &  1659.4  &  1164.5  &  1334  \\ 
ClF  &  1047.6  &  970.8  &  1028.1  &  1158.3  &  1049.8  &  971.4  &  1104  &  855.2  &  1262.2  &  1118.7  &  1019.3  &  1190.6  &  878.6  &  1063  \\ 
HCl  &  727.6  &  686.2  &  733.5  &  720.2  &  664  &  621.4  &  703.9  &  554.7  &  740.9  &  673.1  &  623.1  &  721.6  &  557.5  &  707  \\ 
\multicolumn{15}{c}{} \\
ME  &  -26.4  &  -58.5  &  -45.1  &  74.1  &  9.3  &  -37.8  &  40.4  &  -109.0  &  191.1  &  91.4  &  24.3  &  140.0  &  -69.1  &    \\ 
MAE  &  37.6  &  60.5  &  53.8  &  74.4*  &  21.6*  &  37.8*  &  41.3*  &  109.0  &  191.1  &  98.3  &  52.9  &  140.0  &  69.1*  &    \\ 
\hline
\multicolumn{15}{c}{Overall statistics} \\
ME  &  -11.4  &  -42.9  &  -27.1  &  44.9  &  -6.6  &  -44.2  &  18.9  &  -101.8  &  127.2  &  50.8  &  -0.1  &  89.5  &  -73.4  &    \\ 
MAE  &  28.2  &  44.1  &  41.1  &  49.9*  &  26.0*  &  44.2*  &  31.8*  &  101.8  &  127.8  &  73.0  &  49.6  &  93.6  &  73.4*  &    \\ 
\hline\hline
\end{tabular}
\end{footnotesize}
\end{sidewaystable*}
%

We see that the results depend rather importantly on the used reference
ground-state orbitals.
{This indicates that any non-self-consistent use of the ISI
functional must be considered with the due caution, while only self-consistent
calculations could give definitive information on the real quality of the ISI
energy. However, the self-consistent implementation of the ISI functional is
an extremely hard task. On the other hand, using reference orbitals which} {are
simpler to compute, in order to evaluate ISI functional non-self-consistently
may offer a} {more pragmatic approach that can still provide interesting
information on this method. For this reason we consider this analysis in the following.}

A first inspection of the overall results, i.e. the MAE in the overall statistics 
at the bottom of the table, 
shows that the best ISI results are found 
using PBE0 and LHF orbitals (overall mean absolute errors (MAEs) of 26.0 and 31.8 mHa, respectively).
We remark that these results are of similar quality as those of the
semilocal DFT functionals: the MAE of the best GGA functional (LYP) is 28 mHa.
On the other hand, the use of PBE orbital leads to overestimated 
absolute ISI correlation energies, while the use of HF or BH-LYP orbitals
yields largely underestimated absolute energies.
Similar trends are obtained for the underlying GL2
(MP2 in the case of HF) correlation energies.
It is interesting to see that ISI strongly improves over GL2 for PBE, PBE0 and LHF;
an opposite trend is found for HF, while no relevant differences are found for BH-LYP.

A more detailed analysis of the different systems can be obtained by
inspecting the statistics reported for different classes of systems as well as 
inspecting  Fig.~\ref{fig10} which reports the 
errors on the absolute ISI correlation energies for all the systems.
The plot clearly shows that the use of Hartree-Fock orbitals leads to an underestimation of 
the absolute correlation for all systems. Instead, when LHF and PBE orbitals are considered
atomic correlation energies are computed with quite good accuracy but molecular correlation
energies are significantly overestimated.
This finding has an important effect on the calculation of atomization-correlation energies 
as shown in Table \ref{tab2}.
\begin{table*}
\begin{center}
\begin{scriptsize}
\caption{\label{tab2}Correlation atomization energies (mHa, with opposite sign) from semilocal
 DFT functionals (using PBE orbitals) as well as ISI and GL2 methods calculated using different 
reference ground-state orbitals. Reference data are taken from Ref. \cite{ONeGil-MP-05}. 
The last lines report the mean error (ME) the mean absolute error (MAE) for each case; a star is 
appended to the MAEs to indicate, for each choice of the reference orbitals, the best method between ISI and GL2.}
\begin{tabular}{l|rrr|rrrrr|rrrrr|r}
\hline\hline
Correlation:  & LYP & PBE & PBEloc & \multicolumn{5}{c}{ISI} & \multicolumn{5}{c}{GL2} &\\ 
\cline{2-4}\cline{5-9}\cline{10-14}
Orbitals: &PBE &PBE & PBE & PBE & PBE0 & BHLYP & LHF & HF & PBE & PBE0 & BHLYP & LHF & HF & Ref. \\
\hline
H$_2$  &  38.2  &  42.9  &  37.4  &  38.4  &  35.0  &  32.2  &  36.9  &  28.0  &  53.2  &  46.3  &  41.2  &  50.4  &  34.3  &  41.0  \\ 
NH$_3$  &  126.1  &  134.3  &  134.4  &  189.3  &  171.0  &  157.7  &  182.2  &  135.3  &  246.0  &  213.8  &  192.0  &  233.1  &  158.1  &  152.0  \\ 
CH$_4$  &  136.7  &  155.7  &  152.7  &  170.0  &  157.6  &  146.9  &  166.5  &  128.1  &  212.9  &  192.5  &  175.8  &  207.2  &  148.4  &  143.0  \\ 
H$_2$O  &  83.8  &  89.6  &  90.7  &  165.0  &  147.6  &  135.3  &  156.8  &  115.3  &  211.5  &  182.1  &  162.5  &  197.3  &  133.4  &  116.0  \\ 
FH  &  41.2  &  42.5  &  43.0  &  111.2  &  97.7  &  88.4  &  102.5  &  74.2  &  130.5  &  111.4  &  98.7  &  118.1  &  80.6  &  66.0  \\ 
HCN  &  114.6  &  115.5  &  121.6  &  276.7  &  240.9  &  215.4  &  250.5  &  175.6  &  377.9  &  309.9  &  266.5  &  345.2  &  206.1  &  171.0  \\ 
CO  &  70.3  &  68.9  &  76.4  &  230.5  &  197.9  &  174.4  &  211.3  &  140.3  &  305.7  &  247.9  &  210.3  &  270.5  &  161.0  &  124.0  \\ 
N$_2$  &  100.4  &  91.7  &  99.7  &  280.7  &  242.1  &  215.0  &  267.4  &  172.9  &  386.4  &  313.5  &  267.5  &  359.8  &  203.2  &  173.0  \\ 
C$_2$H$_4$  &  182.1  &  205.1  &  207.2  &  277.3  &  252.3  &  232.2  &  267.6  &  198.5  &  352.1  &  309.8  &  278.4  &  335.7  &  229.4  &  168.0  \\ 
H$_2$CO  &  125.8  &  134.9  &  140.0  &  276.9  &  242.7  &  217.3  &  257.4  &  179.8  &  362.7  &  302.3  &  261.9  &  327.6  &  207.4  &  175.0  \\ 
HOOH  &  123.5  &  128.1  &  135.1  &  316.0  &  275.2  &  246.6  &  301.1  &  202.9  &  416.7  &  344.8  &  298.7  &  388.5  &  234.8  &  201.0  \\ 
F$_2$  &  33.5  &  27.5  &  32.2  &  225.4  &  184.6  &  157.8  &  213.8  &  120.4  &  284.2  &  219.0  &  179.7  &  265.8  &  130.1  &  111.0  \\ 
SiH$_2$  &  69.1  &  69.6  &  65.8  &  94.1  &  79.3  &  69.0  &  87.9  &  57.3  &  97.6  &  78.5  &  65.8  &  79.4  &  61.8  &  62.0  \\ 
PH$_3$  &  110.3  &  116.2  &  113.0  &  152.2  &  133.0  &  117.8  &  144.0  &  97.8  &  166.6  &  140.6  &  110.8  &  150.5  &  96.1  &  112.0  \\ 
SO$_2$  &  116.6  &  116.8  &  131.9  &  476.0  &  391.2  &  333.4  &  418.1  &  260.3  &  606.0  &  471.5  &  377.8  &  510.8  &  279.5  &  221.0  \\ 
ClF  &  36.9  &  33.7  &  39.1  &  189.7  &  153.9  &  129.8  &  170.6  &  100.5  &  205.3  &  160.9  &  130.5  &  183.3  &  89.7  &  76.0  \\ 
HCl  &  37.9  &  41.7  &  41.9  &  80.0  &  71.2  &  64.4  &  80.1  &  55.4  &  82.4  &  73.0  &  63.8  &  83.0  &  56.2  &  43.0  \\ 
&&&  &    &    &    &    &    &    &    &    &    &    &    \\ 
ME  &  -35.8  &  -31.8  &  -29.0  &  82.0  &  54.0  &  34.0  &  68.2  &  5.2  &  137.8  &  91.9  &  60.4  &  114.8  &  20.9  &    \\ 
MAE  &  38.3  &  39.3  &  35.3  &  82.3*  &  54.7*  &  35.1*  &  68.7*  &  12.7*  &  137.8  &  91.9  &  60.5  &  114.8  &  23.6  &    \\ 
\hline\hline
\end{tabular}
\end{scriptsize}
\end{center}
\end{table*}
In this case the smaller errors are found for
HF-based calculations, which benefit from a large error cancellation
effect: indeed, as shown in Fig. \ref{fig10} molecules and atoms are 
underestimated by about the {\it same} quantity.
On the contrary, for all other reference orbitals
an important overestimation of the absolute ISI correlation energy is
observed.
Note that, in any case, the ISI correlation atomization energies
computed for the present test set are always
better than the corresponding GL2 correlation atomization
energies, yielding MAEs of 138.5, 92.5,	62.4, 116.5, and
23.7 mHa for PBE, PBE0, BHLYP, LHF, and HF orbitals respectively.
Moreover, the HF-ISI results are also almost
three times better than those obtainable by semilocal DFT functionals
(the best being PBEloc with a MAE of 36 mHa).

\subsection{Total atomization energies}

In Table \ref{tab2a} we report the total atomization energies.
We compare the ISI results to HF+GGA correlation approaches.
Note that in the latter methods no error cancellation between
exchange and correlation occurs and static-correlation
is not considered \cite{Gri-JCP-97,Han-Mol-01}. 
Thus HF+GGA calculations give much worse results than 
conventional GGA xc approaches.  
However, here they can be used to asses the quality of the ISI results.   
ISI-HF has a MAE of only 11.7 mHa which is 4 times better than HF+GGA.
Conversely, ISI-LHF largely overestimates atomization energies, yielding an absolute 
accuracy close to HF+GGA (which, on the other hand,
underestimate the atomization energies.\cite{ConFabDel-PRB-12})

We note that the present results for ISI atomization energies are
slightly different from the ones reported in the original ISI publication
\cite{SeiPerKur-PRL-00}. This is due to the different 
choice of the parameter $D$ in Eq.~\eqref{eq:WprimePC}, which has been fixed here by 
using the exact value of $W'_\infty[\rho]$ for the He atom density\cite{GorVigSei-JCTC-09} 
instead of the one estimated from a metaGGA functional used in Ref.~\cite{SeiPerKur-PRL-00},
and due to the different basis-set used (recall that in the present work we used
extrapolation towards the complete basis set limit).


\begin{table*}
\begin{center}
\caption{\label{tab2a}Total atomization energies (mHa, with opposite sign) form semilocal DFT functionals (using PBE orbitals) as well as from ISI methods calculated using different reference ground-state densities and orbitals. Reference data are taken from Ref. \cite{ONeGil-MP-05}. 
The last lines report the mean error (ME), the mean absolute error (MAE), and the mean absolute relative error (MARE) for each case.
1mHa=0.62751 kcal/mol}
\begin{tabular}{l|rrr|rrrrr|r}
\hline\hline
Method: & HF+LYP & HF+PBE & HF+PBEloc & \multicolumn{5}{c}{ISI} \\ 
\cline{5-9}
orbitals: & PBE& PBE & PBE& PBE & PBE0 & BHLYP & LHF & HF & Ref. \\
\hline
H$_2$   & 172.3  & 164.5  & 163.8  & 171.2  & 168.2  & 165.5  & 170.5  & 161.6  & 174.5  \\
NH$_3$  & 446.4  & 436.6  & 443.9  & 506.1  & 489.8  & 476.8  & 500.4  & 455.5  & 475.5  \\
CH$_4$  & 660  & 655  & 661.8  & 689.3  & 678.7  & 668.9  & 687.4  & 651.4  & 626        \\
H$_2$O  & 332.1  & 325.9  & 331.9  & 411  & 395  & 383  & 404  & 363.6  & 371            \\
HF      & 195.6  & 190.9  & 193.9  & 264.2  & 251.6  & 242.6  & 256.6  & 228.7  & 216.4   \\
HCN     & 431.8  & 426.7  & 435.2  & 586.1  & 554  & 530.4  & 459.3  & 492.7  & 496.9     \\
CO      & 348.5  & 347  & 354.6  & 500.9  & 471.8  & 450.9  & 486  & 418.5  & 413.8     \\
N$_2$   & 284.3  & 275.6  & 283.6  & 457.4  & 422.2  & 396.9  & 446.5  & 356.7  & 363.7   \\
C$_2$H$_4$  & 864.6  & 863.6  & 875.6  & 951.6  & 930.3  & 912.4  & 945.7  & 881  & 898.8 \\
H$_2$CO  & 536  & 533.1  & 543.1  & 676  & 646.9  & 625  & 662.4  & 590  & 596.7          \\
HOOH  & 333.9  & 326.7  & 338.4  & 516.7  & 481  & 454.8  & 505  & 413.4  & 428.9          \\
F$_2$  & 25.8  & 31.8  & 27.1  & 156.9  & 120.3  & 96.5  & 146.9  & 61.1  & 62.5    \\
SiH$_2$  & 245.5  & 233.9  & 235.1  & 265.1  & 252.7  & 244.4  & 261.7  & 233.6  & 242.9  \\
PH$_3$  & 385.2  & 373.1  & 377.2  & 418.2  & 402.8  & 390.5  & 414.2  & 372.7  & 387.2   \\
SO$_2$  & 290.1  & 290.4  & 305.4  & 624.6  & 551.3  & 502.3  & 580.7  & 433.9  & 414.2   \\
ClF  & 50.5  & 47.3  & 52.6  & 193.8  & 162.4  & 141.8  & 178.3  & 114  & 100.1           \\
HCl  & 160.6  & 158.4  & 161  & 202.4  & 194  & 187.3  & 202.3  & 178.1  & 171.2  \\
 & & & & & & & & & \\
ME &  -42.9 & -48.4 & -41.8 & +67.7 & +43.1 & +25.3 & +51.0 & -2.0 & \\
MAE & 47.2 & 51.8 & 46.0 & 68.1 & 43.8 & 26.3 & 55.9 & 11.7 & \\
\hline\hline
\end{tabular}
\end{center}
\end{table*}

\subsection{Main-group chemistry benchmark}\label{sec:thermo}
To assess the practical applicability of the ISI functional to main-group
chemistry, we have performed a series of tests involving different properties
of interest for computational chemistry. 
We have restricted our study to ISI calculations
employing HF and LHF reference orbitals
(hereafter denotes as ISI-HF and ISI-LHF, respectively).
This choice was based on the fact that, as explained in
subsection 
\ref{refsect}, ISI-HF is expected to yield
the best performance for these tests (according
to the results of Table \ref{tab2}) while ISI-LHF 
{provides the best} {approximation for the}
performance of self-consistent ISI 
calculations.
For comparison, we report also the MP2 and B2PLYP \cite{Gri-JCP-06} results,
which are based on GL2 energies, as well as the performance of 
calculations using the popular PBE functional \cite{PerBurErn-PRL-96} and of the Hartree-Fock exchange 
coupled with the semilocal PBEloc correlation \cite{ConFabDel-PRB-12}
 (HF+PBEloc). The latter is a simple approach
adding semilocal dynamical correlation to Hartree-Fock and can
give information on the possible accuracy of 
``standard'' DFT methods when used together with exact exchange;
we remark anyway that much improved results can be obtained by
more sophisticated DFT approaches including static and/or strong
correlation treatments \cite{Bec-JCP-05,Bec-JCP-13a}.

In the upper part of Table \ref{tab_mae1} we report 
the mean absolute errors (MAEs) for several standard tests 
concerning thermochemical properties.
\begin{table*}
\begin{center}
\caption{\label{tab_mae1} Mean absolute errors (kcal/mol) on several tests as obtained from ISI calculations using LHF and HF orbitals. 
PBE, HF+PBEloc, MP2 and B2PLYP results are reported for comparison. The best (worst) result for each test is in boldface (underlined).
The last line reports the relative MAE with respect MP2 (see Eq. (\ref{ermae})).}
\begin{tabular}{lrrrrrr}
\hline\hline
Test & PBE & HF+PBEloc & MP2 & B2PLYP & ISI-HF & ISI-LHF \\
\hline
\multicolumn{7}{c}{Thermochemistry} \\
AE6   & 13.3 & 24.0 & 9.6 & {\bf 1.6} & 10.0 & \underline{43.4} \\
G2/97 & 14.7 & 26.3 & 12.3& {\bf 4.0} & 15.9 & \underline{53.1}  \\
IP13  & 3.3 & \underline{7.0} & 2.2 & {\bf 1.9} & 3.0 & 6.0 \\
EA13  & {\bf 2.8} & 9.0 & 3.4 & 4.1 & 5.9 & \underline{9.3} \\
PA12  & 2.2 & \underline{6.6} & 1.0 & 1.4 & {\bf 0.9} & 2.6 \\
K9    & 7.4 & 4.3 & 4.1 & {\bf 1.6} &  7.2 &  \underline{8.5} \\
BH76  & 9.7 & 6.8 & 5.2 & {\bf 2.2} &  10.1 & \underline{11.7}  \\
BH76RC & 4.3 & 6.9 & 3.9 & {\bf 1.2} &  7.0 & \underline{16.4} \\
\multicolumn{7}{c}{Non-covalent interactions} \\
HB6   & {\bf 0.4} & \underline{1.7} & {\bf 0.4} & {\bf 0.4} &  0.7 & 1.1 \\
DI6   & {\bf 0.4} & {\bf 0.4} & 0.5 & 0.5 &  0.8 & \underline{3.3} \\
CT7   & 2.3 & 1.1 & 0.8 & {\bf 0.6} &  2.2 & \underline{7.5} \\
DHB23 & 1.0 & 1.0 & 1.3 & {\bf 0.5} &  5.1 & \underline{11.0} \\
S22   & \underline{2.7} & 1.9 & 1.2 & 1.9 &  {\bf 0.4} & 1.5 \\
\multicolumn{7}{c}{Statistics} \\
RMAE  & 1.50 & 2.31 & 1.00 & {\bf 0.75} & 1.71 & \underline{4.14} \\
\hline\hline
\end{tabular}
\end{center}
\end{table*}
In the last line of Table \ref{tab_mae1} we report, for each method $X$, the relative mean absolute error 
(RMAE) with respect MP2, i.e.
\begin{equation}\label{ermae}
RMAE^{X}=\sum_i \frac{MAE_i^{X}}{MAE_i^{MP2}}\ ,
\end{equation}
where $i$ indicates the different tests. 
 
The results clearly show that ISI-LHF often gives the largest MAEs, 
with a RMAE of 4.1.
Significantly better results are obtained by ISI-HF calculations (RMAE=1.7).
However, the performance for barrier heights
(BH76 and K9) is quite poor and even worse
than that obtained by adding a simple semilocal
correlation to Hartree-Fock.\cite{ConFabDel-PRB-12,FabTreTerCon-JCTC-14}
We note also that for this property Hartree-Fock
and LHF based ISI calculations yield a
quite similar performance.
On the other hand,  ISI-HF yields the best results for the PA13 test and the S22 test.
%


When the focus is on non-covalent interactions
(bottom part of Table \ref{tab_mae1}),
ISI-HF performs quite
well for both hydrogen bond (HB6) and
dipole-dipole (DI6) interactions having a comparable accuracy as
MP2 and B2PLYP.
The ISI-HF functional
outperforms other approximations for the S22 test,
which contains different kinds of biology relevant
non-covalent complexes having hydrogen-bond, dipole-dipole,
and dispersion character.

\subsection{A closer look at dispersion complexes}

\begin{figure}[t]
\includegraphics[width=0.9\columnwidth]{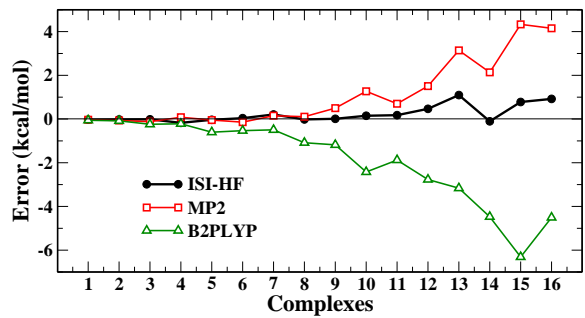}
\caption{\label{fig_disp} Signed errors (kcal/mol) in the calculation of the interaction energy 
of different dispersion complexes (1:He-Ne, 2:Ne-Ne, 3:CH$_4$-Ne, 4:CH$_4$-F$_2$, 5:CH$_4$-CH$_4$, 
6:C$_6$H$_6$-Ne, 7:CH$_2$-CH$_2$, 8:C$_2$H$_4$-C$_2$H$_4$, 9:C$_6$H$_6$-CH$_4$, 10:C$_6$H$_6$ sandwich dimer, 
11:C$_6$H$_6$ T-shaped dimer, 12:C$_6$H$_6$ displaced dimer, 13:Pyrazine-dimer, 14:Uracil stacked dimer, 
15:Adenine-Thymine stacked dimer, 16: Indole-Benzene stacked dimer).}
\end{figure}

The small error for the S22 test set suggests that  ISI-HF may be
more accurate than other approaches (e.g. B2PLYP) in
the description of dispersion complexes.
As further evidence, we report in Fig.~\ref{fig_disp}
the signed error obtained from ISI-HF, MP2, and
B2PLYP in the calculation of the interaction energy
of a collection of different dispersion complexes,
which includes the dispersion-dominated S22 cases as well as additional
test cases from the literature.\cite{RezRilHob-JCTC-11,RezRilHob-JCTC-12,GraFabDel-PCCP-13,FabDelGra-CPL-15} 
It can be seen that, indeed, ISI-HF results are always
very accurate ($\lessapprox$1 kcal/mol), whereas for the dimers of aromatic molecules  
MP2 (B2PLYP) largely overestimate (underestimate) the interaction energy.

\begin{figure}[t]
 \includegraphics[width=0.96\columnwidth]{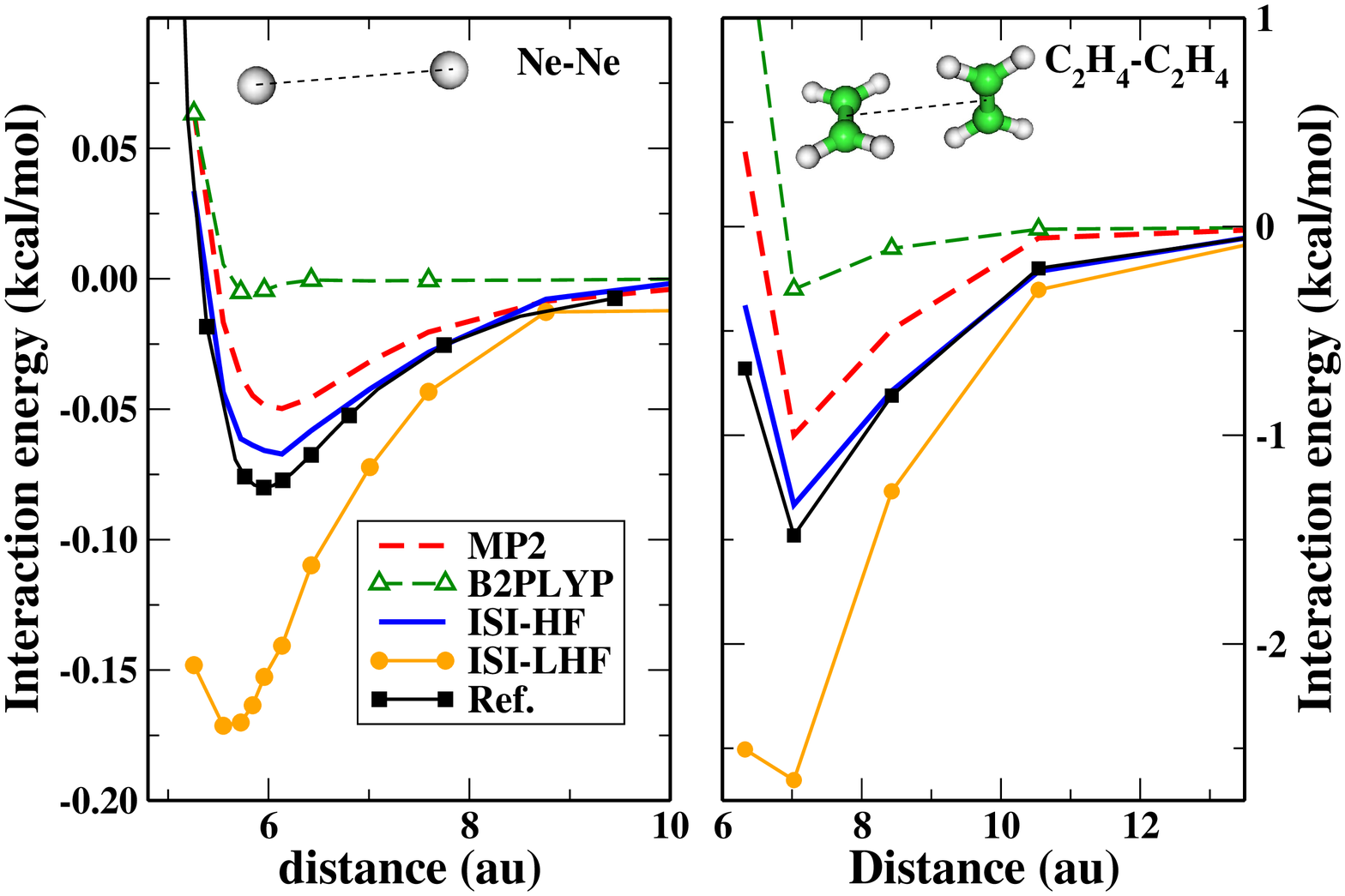}
 \caption{\label{fig:disscurves} Interaction energy curves  for Ne-Ne and C$_2$H$_4$-C$_2$H$_4$. All energies have been corrected for the basis-set superposition error. Reference values for Ne-Ne and C$_2$H$_4$-C$_2$H$_4$ have been taken from Refs. \cite{FabDelGra-CPL-15} and \cite{grafova2010}, respectively.}
\end{figure}

In Fig.~\ref{fig:disscurves} we also report the interaction energy curves 
for Ne-Ne and C$_2$H$_4$-C$_2$H$_4$, which show again that ISI-HF accurately captures dispersion interactions. {Further analysis and discussion of these results are reported in subsection \ref{subsec:disp}.}


\subsection{Static correlation}
One of the purposes of including the strong-coupling limit into approximate functionals is the hope to capture
 static correlation without resorting to symmetry breaking. However, it is already clear from Eq.~\eqref{eq:EcGL2mininf} that 
the ISI functional will not dissociate correctly a single or multiple bond in a restricted framework. In fact, as the bond is
 stretched, the ISI xc energy of Eq.~\eqref{eq:EcGL2mininf} will be quite different than the one for the two equal open shell fragments. 
The problem is that only the electrons involved in the bonds should be strongly correlated. The rest of the fragment should
 be in the usual weak or intermediate correlation regime, but the global interpolation makes the whole fragment be in the
 strong-coupling regime. A local interpolation might fix this issue, but it needs to be constructed carefully.\cite{VucIroSavTeaGor-JCTC-16}

An exception is the H$_2$ molecule for which all the electrons are involved in the bond. Indeed, Teale, Coriani and Helgaker\cite{TeaCorHel-JCP-10} 
had found a very good agreement between the ISI model for the adiabatic connection curve (in a restricted framework) 
and their accurate results in the case of the H$_2$ molecule dissociation, when the bond is stretched up to 10 bohr.  
Their study used full configuration-interaction 
(FCI) densities and the corresponding KS orbitals and orbital energies from the Lieb maximization procedure as input quantities. 
They have also tested how the choice of the parameter $D$ in Eq.~\eqref{eq:WprimePC} affects the shape of the adiabatic connection curve. 
They found that the original metaGGA choice used in Ref.~\cite{SeiPerKur-PRL-00} does not yield accurate results. 
whereas the parameter $D$ used here was found to yield rather accurate results up to 10 bohr.

\begin{figure}[t]
\includegraphics[width=\columnwidth]{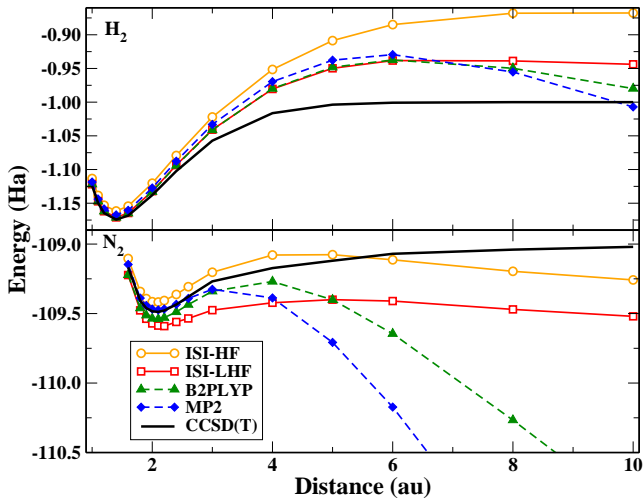}
\caption{\label{fig:static} Dissociation energy curves for the H$_2$ and N$_2$ molecules in 
a spin-restricted formalism. 
In this case second-order perturbation theory diverges as the molecule is stretched, and the ISI functional tends to Eq.~\eqref{eq:EcGL2mininf}.
Note also that ISI-LHF and ISI-HF will coincide at infinite distance.
}
\end{figure}

In Fig.~\ref{fig:static} we report the dissociation curves of the H$_2$ and N$_2$ molecules in a spin-restricted formalism for different methods.
Our ISI results are not very accurate if compared to the reference CCSD(T) results, but qualitatively better than MP2 and B2PLYP which diverge for large distance.
The inaccuracy of our ISI results  originates form the approximated LHF (or HF) densities, orbitals and orbital energies: in fact, the spin-restricted 
ISI turns out to be very sensitive to the input ingredients. The ISI results in Ref. \cite{TeaCorHel-JCP-10} are much more accurate due to the fact 
that FCI input density, orbitals and orbital energies have been used.
Moreover for H$_2$, we recall that the ISI results will be exact at infinite distance only if the parameter $D$ is self-interaction free for the 
H atom density.\cite{SeiPerKur-PRA-00} 

\section{Discussion}
The results reported in 
Section~\ref{sec:results} 
show that 
the performance of ISI-HF is quite good when compared with 
HF+GGA methods (eg. HF+PBEloc), since the former describes 
dynamical and static correlation without any error cancellation
while the latter do not. 
On the other hand ISI-HF is much less appealing, if compared  
to MP2 which yields in many cases better results at similar computational cost. 
One important exception are dispersion interactions, for which ISI-HF outperforms MP2. 
Instead when ISI is applied to DFT orbitals (i.e. LHF) the results are 
rather bad. 
In the following subsections we try to analyze and rationalize this performance, in order
to provide useful information which can be used to improve functionals based on
interpolations between the weak and the strong interacting limits.

\subsection{Influence of the size-consistency problem}
\label{sec:sizecons}
Being a non-linear function of exact exchange and GL2 total energies,
the ISI xc energy functional is formally not size consistent. This means that
computing the (spin-unrestricted or spin-restricted) ISI xc energy of two 
systems separated by a distance large
enough (eventually infinite) to make the interaction between 
them negligible, yields a  result which is different form the sum of the 
ISI xc energies of the two isolated 
systems. 

One exception is the case of a set of identical systems,
e.g. a homonuclear dimer $A-A$, where $A$ is closed-shell or the spin-unrestricted formalism is used:  
under these conditions  $E_x$, $E_{\rm GL2}$, $W_\infty$, and $W'_\infty$ are all size-consistent, thus
$X[A-A]=2X[A]$, while $Y[A-A]=Y[A]$ and $Z[A-A]=Z[A]$. Since the ISI xc energy (see Eq. (\ref{eq:isif})) is linear in $X$
and $W_\infty$ is a size-consistent quantity, the the whole result is size-consistent.

The issue of size-inconsistency may, of course, 
affect the results when atomization or
interaction energies are calculated.
To investigate the relevance of this problem, we perform
a numerical study on the magnitude of this effect.
Consider a system $M$ (e.g., a molecule) composed of different fragments $A_i$ (e.g. atoms) with $i=1,\ldots,N$. 
The total xc interaction energy in this system is
\begin{equation}\label{e3}
E_{xc}^{int}\left(M\right) = E_{xc}(M)-\sum_{i=1}^{N}E_{xc}(A_i).
\end{equation}
Here, $E_{xc}(M)$ denotes the xc energy of $M$ and $E_{xc}(A_i)$ the xc energy of the {\em isolated} fragment $A_i$.
Consequently, if we denote with $M^*$ the system
obtained by bringing all fragments $A_i$ at large distance from each other
(such that their mutual interaction is negligible), this
interaction energy can also be written as
\begin{equation}\label{e4}
E_{xc}^{int}\left(M\right) = E_{xc}\left(M\right)-E_{xc}\left(M^*\right).
\end{equation}
For any size-consistent method, Eqs. (\ref{e3}) and (\ref{e4}) give the
same result. However, for a non-size-consistent method such as ISI, their
difference
\begin{equation}\label{e5}
\Delta_{xc}\left(M\right) = E_{xc}\left(M^*\right)-\sum_{i=1}^{N}E_{xc}(A_i)
\end{equation}
can provide a measure for the size-consistency problem
(clearly, $\Delta_{xc}=0$ for any size-consistent method).

In the specific case of ISI we have
\begin{eqnarray}\label{e6}
&& E_{xc}^{\rm ISI}\left(M^*\right)=\\
\nonumber
&& = f^{\rm ISI}\Big(E_x\left(M^*\right),E_{\rm GL2}\left(M^*\right),W_\infty\left(M^*\right),W'_\infty\left(M^*\right)\Big),
\end{eqnarray}
where $f^{\rm ISI}(w_1,w_2,w_3,w_4)$ is the non-linear function
of four variables defined in Eqs.~\eqref{eq:isif}
 and \eqref{eq:ISIxyz}.
Assuming that all four ingredients are size-consistent, we can
further write
\begin{eqnarray}
\nonumber
E_{xc}^{\rm ISI}\left(M^*\right) &= & f^{\rm ISI}\Bigg(\sum_{i=1}^NE_x\left(A_i\right),\sum_{i=1}^NE_{\rm GL2}\left(A_i\right),\\
\label{e7}
&&\quad \sum_{i=1}^NW_\infty\left(A_i\right),\sum_{i=1}^NW_\infty'\left(A_i\right)\Bigg)\ .
\end{eqnarray}
Even then, we typically have $\Delta_{xc}^{\rm ISI}(M)\ne0$, since $f^{\rm ISI}$ is not linear, i.e.
\begin{equation}
E_{xc}^{\rm ISI}(M^*)\ne\sum_{i=1}^NE_{xc}^{\rm ISI}(A_i).
\end{equation}
As previously mentioned, an exception arises in cases with identical fragments, $A_i=A$ (all $i$), 
since the function $f^{\rm ISI}$ has the property
\begin{equation}
f^{\rm ISI}(Nw_1,Nw_2,Nw_3,Nw_4)=Nf^{\rm ISI}(w_1,w_2,w_3,w_4).
\nonumber\end{equation}

Using Eq. (\ref{e6}) and the corresponding expression for $E_{xc}^{\rm ISI}(A_i)$, it is
possible to evaluate the effect of the size-consistency violation
of ISI for different systems.
The results of these calculations are reported, 
for a selected test set of molecules, in Table \ref{tab4}.
In these calculations we have considered
a spin-unrestricted formalism for HF and GL2 calculations
on open-shell atoms, assuming
that the corresponding results are properly size-consistent
(whether this is formally correct is still under debate in literature
\cite{Sav-CP-09}; however, numerical results suggest that our approximation is
quite accurate in the considered cases). 
\begin{table}[t]
\begin{center}
\caption{\label{tab4}Values of $\Delta_{xc}$ per bond (in mHa), calculated using  Eqs. (\ref{e5}), (\ref{e6}), and (\ref{e7}), 
for a selection of molecular systems. Note that since $E_x$ is a size-consistent quantity $\Delta_{xc}=\Delta_c$.}
\begin{tabular}{lrclr}
\hline\hline
Molecule & $\Delta_{xc}$ & $\;\;\;$ & Molecule & $\Delta_{xc}$ \\
\hline
%
CH$_4$ & 0.03 & & H$_2$CO & 0.01 \\
NH$_3$ & 0.03 & & HOOH & 0.02 \\
H$_2$O & 0.04 & & SiH$_2$ & 0.00 \\
FH & 0.04 & & PH$_3$ & 0.00 \\
HCN & 0.01 & & SO$_2$ & -2.44 \\
CO & -0.02 & & HCl & 0.00 \\
C$_2$H$_4$ & 0.02 & & PN & -1.39 \\
SiC & -1.70 & & SiO & -3.43 \\
PO & -3.25 & & NCl$_3$ & -1.33 \\
\hline\hline
\end{tabular}
\end{center}
\end{table}
Inspection of the Table shows that for molecules
composed of first row elements (plus hydrogen)
the values of $\Delta_{xc}$ are negligible.
Thus, the ISI functional behaves, in practice,
as a size-consistent method. On the other hand,
for molecules including both first and second row
elements, larger values are found. We remark 
that these values are, anyway, often smaller than few
mHa per bond, so that the size-inconsistency problem is
not too large also in these cases.

\begin{figure}[t]
\includegraphics[width=0.9\columnwidth]{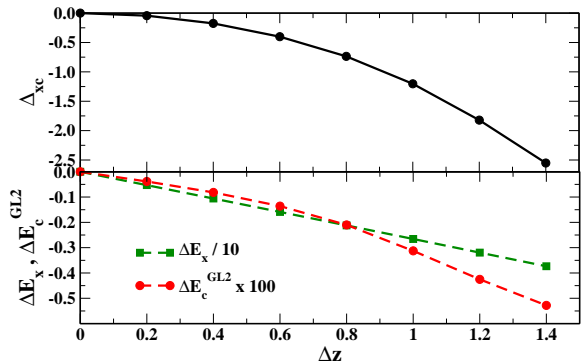}
\caption{\label{fig3}Upper panel: Values of $\Delta_{xc}$ as functions of $\Delta Z$
 for the dissociation of N$_2$ into two atomic fragments having 7 electrons and nuclear 
charges $Z_1=7+\Delta Z$ and $Z_2=7-\Delta Z$. Lower panel: Values of $\Delta E_x=E_x[fragment1]-E_x[fragment2]$ 
and $\Delta E_c^{\rm GL2}=E_c^{\rm GL2}[fragment1]-E_c^{\rm GL2}[fragment2]$ as functions of $\Delta Z$. The values of $\Delta E_x$ and $\Delta E_c^{\rm GL2}$ have 
been scaled only for graphical reasons.}
\end{figure}

The difference between the two kinds of behaviors
observed in Table \ref{tab4} traces back to 
the fact that when only first row elements are present,
all atoms display quite similar values of exchange
and GL2 correlation; thus, the ISI behavior is rather similar to 
the ideal case of identical systems and the 
size-consistency violation is small.
On the contrary, when both first and second row atoms
are present, the atomic properties are
significantly different and the non-linear nature of the
ISI formula leads to a non-negligible size-inconsistency.
Further evidence of this fact is given in Fig. \ref{fig3},
where we report the values of $\Delta_{xc}$ for the
atomization of a N$_2$ molecule into two atomic fragments
having 7 electrons each (as the N atom) but
nuclear charges $Z_1=7+\Delta Z$ and
$Z_2=7-\Delta Z$, for various values of $\Delta Z$.
Indeed, the plot clearly shows that the size-consistency problem
grows with the difference between the two atomic fragments.

\subsection{Role of the energy-gap}
\label{subsec:gap}
The fact that similar trends are observed in Tables \ref{tab1} and \ref{tab2} 
for ISI and GL2 correlation energies for
different reference orbitals and different systems, suggests that
the energy gap between occupied and unoccupied molecular orbitals may play 
a major role in determining the accuracy of the ISI 
correlation energy. This difference
is in fact smaller for semilocal DFT (PBE) and larger for HF, having
intermediate values for hybrid and the LHF methods.
Similarly, the energy gap is larger for closed shell atoms
and smaller for open-shell atoms and molecules.
These observations fit well with the behavior reported in Table 
\ref{tab1} and Fig. \ref{fig10}.

To investigate this feature, we have considered for
all the systems in Table \ref{tab1} the application
to the LHF ground-state orbitals of a {\it scissor operator} to rigidly
move all the unoccupied orbitals up in energy by
\begin{equation}\label{escissor}
\Delta E = \alpha\left(E_{g}[HF]-E_{g}[LHF]\right)\ ,
\end{equation}
where $E_{g}[HF]$ and $E_{g}[LHF]$ are the HOMO-LUMO
gaps for HF and LHF, respectively, while $\alpha$ is a
parameter used to tune the effect.
Thus, for $\alpha=0$ no shift is applied, whereas for 
$\alpha=1$ the applied shift is such that the LHF
HOMO-LUMO gap is lifted up to the HF value.

\begin{figure}[t]
\begin{center}
\includegraphics[width=1.0\columnwidth]{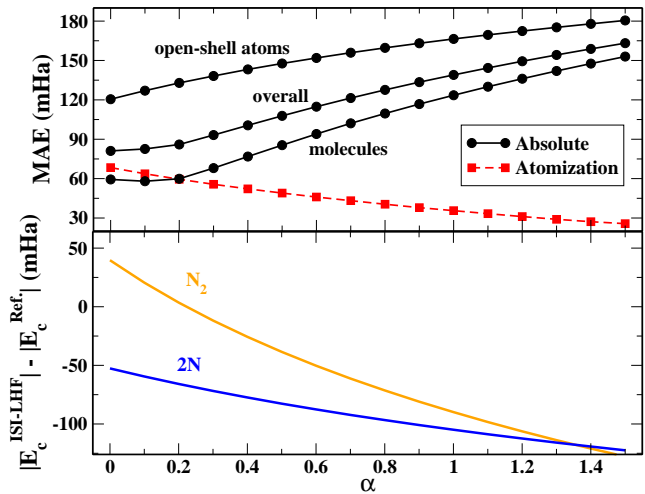}
\caption{\label{fig_scissor}Top panel: Mean absolute errors (MAEs) for the ISI correlation and atomization correlation energies of the systems 
of Table \ref{tab1} as functions of the $\alpha$ parameter of Eq. (\ref{escissor}).
 Bottom panel: Deviations from reference values of the ISI 
correlation energies of the N$_2$ molecule and two N atoms as functions of the $\alpha$ parameter 
of Eq. (\ref{escissor}). All results were computed using a cc-pV5Z basis set.}
\end{center}
\end{figure}

In the bottom panel of Fig. \ref{fig_scissor}
we report the deviations from reference values of 
the ISI correlation energies of the N$_2$ molecule and twice the 
N atom, as functions of the $\alpha$ parameter of Eq. (\ref{escissor}).
The atomization correlation energy error is thus the difference between these two curves.
For simplicity we considered here results with the 5Z basis set and not the extrapolated CBS ones. 
Hence, even at $\alpha=0$ this results for N$_2$ are slightly different from the ones
reported in Tables \ref{tab1} and \ref{tab2}.
At $\alpha=0$ we have an atomization correlation energy error of about 100 mHa.
When $\alpha$ is increased the absolute correlation energies decreases due to an increased 
energy in the denominator.
However, the slopes of the lines are different.
At $\alpha\approx1$ (i.e at the HF gap) the two lines almost cross, meaning that 
the ISI-HF method yields the correct atomization correlation energy.

The MAEs computed for different values of the parameter $\alpha$
for the ISI correlation energies of open-shell atoms, molecules and both,
as well as the MAE of the correlation atomization energies, are reported in the
upper panel of Fig. \ref{fig_scissor}.
The plot shows that the application of a shift for the
unoccupied orbitals generally leads to a worsening of the
ISI correlation energies. This is particularly true
for atoms, which already suffer for an underestimation 
(in absolute value) of the correlation energy, thus increasing the
energy difference between occupied and virtual orbitals 
adds a further underestimation. For molecules instead
at low values of $\alpha$ a moderate improvement
of the correlation energy is observed, since in most molecules
LHF-ISI overestimates (in absolute value)
correlation energies that are thus improved by the 
application of a shift. Nevertheless, for larger values of 
$\alpha$ in all molecules an underestimation
of the correlation is found, so the results rapidly worsen with increasing
shift. We note that the rate of worsening for molecular
correlation energies is quite faster that that observed 
for atoms. 

\subsection{Dispersion interactions}
\label{subsec:disp}
The good performance of the ISI-HF for dispersion interactions is surprising 
and deserves further thoughts. First of all, we notice that the functional $E_{xc}^{\rm ISI}[\rho]$ defined
 by Eqs.~\eqref{eq:isif}-\eqref{eq:WprimePC} inherits (at least for the case of equal fragments)
 the long-range $\sim R^{-6}$ dispersion interaction energy dependence from its $E_{\rm GL2}$ component 
(MP2 in the case of HF reference orbitals considered here). Yet, it systematically outperforms MP2, 
suggesting that it adds a sensible correction to it. The analysis of Str\o msheim {\it et al.}\cite{StrKumCorSagTeaHel-JCP-11} 
shows that the adiabatic connection curve for the interaction energy of dispersion complexes deviates significantly 
from the linear behavior, requiring a considerable amount of ``non-dynamical'' correlation, which seems to be well 
accounted for by the ISI functional (although the picture may be different with HF orbitals).

A possible explanation may be derived by looking at the functional $W_\infty'[\rho]$, which describes the 
physics of coupled oscillations of localized electrons. Its PC semilocal approximation of Eq.~\eqref{eq:WprimePC} 
is a quantitatively good approximation of this energy.\cite{GorVigSei-JCTC-09} The physics of dispersion interactions 
is actually very similar, describing oscillations of coupled charge fluctuations on the two fragments. We suspect 
that when looking at the interaction energy, the physics introduced by $W_\infty'[\rho]$ (when subtracting the internal part of each fragment)
 is actually correct. 
However a more detailed study of this aspect is required and it will be the object of our future work.

\section{Conclusions and Perspectives}
 We have reported the first detailed study of the performances of a functional that includes (in an approximate way) 
the strong-coupling limit, analysing its dependence on basis-set, reference orbitals and other aspects such as the size consistency 
error. Overall, the ISI functional has serious limitations, which could have been expected from some of its formal deficiencies. 
We have rationalized our findings, providing useful information for functionals that can retain the information from the strong-coupling 
limit while remedying to these deficiencies.\cite{VucIroSavTeaGor-JCTC-16} 
In future work, we plan to extend our analysis to functionals 
based on {\rm local} interpolations along the adiabatic connection,\cite{ZhoBahErn-JCP-15,VucIroSavTeaGor-JCTC-16,KonPro-JCTC-16,Bec-JCP-13a,Bec-JCP-13b,Bec-JCP-13c} 
implementing the needed input quantities.

An unexpected finding that emerged from our study is a very good performance of the ISI functional (when used as a correction to Hartree Fock)
 for dispersion interactions, yielding a mean absolute error of only 0.4~kcal/mol on the S22 set, and consistently improving over MP2 in a
 significative way for dispersion complexes (see Figs.~\ref{fig_disp}-\ref{fig:disscurves}). We suspect that the functional $W'_\infty[\rho]$,
 which describes coupled oscillations of localized electrons, is able to capture the physics of interaction energy in dispersion complexes. 
This is an interesting perspective for the ISI functional, which we will investigate in detail in a future work.

\subsection*{Acknowledgments}
We thank Evert Jan Baerends, Oleg Gritsenko and Stefan Vuckovic for insightful discussions and TURBOMOLE GmbH for providing the TURBOMOLE program.
PG-G and MS acknowledge the European Research Council under H2020/ERC Consolidator Grant ``corr-DFT'' [grant number 648932] for financial support.

 \bibliography{biblioPaola.bib,biblio_spec.bib,biblio_add.bib}

\end{document}